# Indium Tin Oxide/Barium Strontium Titanate THz Sensor Antenna

Sai Dittakavi
*Department of Electrical Engineering*
*Northern Illinois University*
Dekalb, USA
sdittakavi@niu.edu

Ibrahim Abdel-Motaleb
*Department of Electrical Engineering*
*Northern Illinois University*
Dekalb, USA
ibrahim@niu.edu

***Abstract*— A Bow-Tie ITO/BST and Au/BST antennas were designed and analyzed using the finite element, Multiphysics COMSOLTM program. The study shows that the outputs peak at several modes and ITO antennas allow for the integration of the antenna with optoelectronic devices.**

***Keywords—THz. Nano-Antenna, COMSOL FEM Modelling***

## I. INTRODUCTION (*HEADING 1*)

Terahertz radiation gap is normally defined as the electromagnetic spectrum that occupies the range from 0.3 THz – 3 THz. This range is also known as the sub-mm waves, since its wavelength falls between 1 mm and 0.1 mm [1]. This wavelength region is sandwiched between the micro-wave and infra-red regions. THz spectrum offers many applications in engineering such as medical imaging, security, communications, scientific instrumentations, and military applications [2]. One of the main advantages of THz radiation is its non-ionizing property. This property makes THz safe for use in medical application, where tissues and DNA will be safe from damage.

The antennas used in this study are Bow-Tie shape metal deposited on top of dielectric substrates. The performance of the antenna is affected by the dielectric strength of the substrate. Low dielectric substrates, such as silica, have been used to build THz antennas, and their performance have been reported in the literatures.

The metal used to build THz antennas is normally gold (Au). Gold is chosen because it is chemically inert and has a very high conductivity. However, Au is opaque to in the optical spectrum; therefore, the ability to achieve vertical integration with optical/optoelectronic devices may be limited. Using a transparent metal may provide the advantage of vertical integration with optical devices. Indium-Tin-Oxide (ITO) can provide this advantage, if used as a metal for the antenna. ITO is a transparent wide bandgap degenerate semiconductor with very high conductivity, where it can act as a metal.

In this study, a Bow-Tie Nano-antenna made of ITO on Barium Strontium Titanate (BST) is designed and analyzed using finite element numerical analysis in the THz range. BST materials can have dielectric constants as high as 300, if not higher. For comparison, a similar antenna using Au on BST is designed and analyzed using the same numerical analysis. The wide bandgaps, of both the BST and the ITO, allow almost all infrared and visible light photons to be transmitted through. To the best of our knowledge, there is no similar THz antenna built using ITO on BST published.

## II. ANALYSIS

COMSOL$^{TM}$ Multi-physics program is used to conduct the study. Bow-Tie antenna shape has been studied by many research groups [3-5]. They have shown that the antenna's performance is enhanced when this Bow-Tie shape is used. This antenna is basically a two triangles with a Nano-gap between their vertices, as shown in Fig. 1. The antenna has 16 μm length, 9 μm width, and 1μm thickness. The gap is 100 nm between the two apexes.

THz antennas can be built using many geometries, such as spiral or rectangular shapes. However, the Bow-Tie structure can provide high electric field, since the two sides of the antenna forms a gap between two sharp edges. Since charges accumulate at sharp edges, the density of the charges at the vertices will be extremely high. And since the electric field is proportional to the charge density, it is expected to have high field intensity for the Bow-Tie antenna. Using arrays from these antennas, the output voltage and current from the arrays are expected to be reasonably high.

The simulation geometry with the various boundaries used in COMSOL simulation is shown in Fig. 2. The antenna is placed on top of 20 μm BST substrate. On the top of the antenna, a 20 μm of air is placed. To avoid reflections from the top boundaries, a Perfectly Matched Layer (PML) with thickness of 10 μm is used on the top and under the bottom of the model [5]; see Fig. 2. The antenna is illuminated vertically by an Electromagnetic wave using a COMSOL feature port that applied 1 V/m magnitude in the z-direction. The port is excited with an electric source placed at 20 μm above the antenna metal, just below the upper PML layer.

The dielectric constant is obtained from the Drude model [5].

$$\varepsilon(\omega) = \varepsilon_\infty - \omega_p^2/(\omega^2 - j\omega\omega_\tau) \quad (1)$$

where, $\varepsilon_\infty$ represents the contribution of the bound electrons to the relative dielectric constant, $\omega_p$ is the plasma frequency, and $\omega_\tau$ is the damping frequency. The dielectric parameter is a complex value, since it is frequency dependent, as shown in (1). COMSOL uses

(1) to obtain the value of the complex dielectric at various frequencies [6].

As mentioned before, to eliminate reflection, PMLs are used. They are not boundary condition but an additional domain that absorbs incident wave without producing reflections. The captured electric field at the feeding gap is the output of the Nano-antenna. In COMSOL, Maxwell equations used to obtain the electric field is equation (2)

$$\nabla * \mu_r^{-1}(\nabla * E) - k_0^2 \varepsilon_r E = 0 \qquad (2)$$

where, $\mu_r$ is the relative permeability, and $\varepsilon_r$ is the relative dielectric constant, $k_0$ is the wave number which is = (ω/c), ω is the radial frequency, and c is the speed of light.

The gap between the two triangles of the Bow-Tie dipole antenna affect the performance of the antenna. The electric field is typically concentrated at gap of the antenna due to Coulomb field. In this case, the gap acts as a capacitor [7]. The effect of gap dimension on the antenna's performance is investigated to obtain the optimum dimension.

## III. RESULTS AND DISCUSSIONS

To ensure that our simulation is correct, a Bow-Tie Au/Silica glass antenna was simulated and compared with the published results reported in [5]. The dimensions used for both antennas are those shown in Fig. 1. In this simulation the dielectric constant for silica is taken to be 2.03, the conductivity of Au is taken to be $45.6 \times 10^6$ S/m, and the complex dielectric constant to be -8.49+1.62j. The published results are shown in Fig. 3 and our results are shown in Fig. 4. As can be seen from the figures, our simulation results are almost identical to the published results. The match between the two results provided the necessary confidence in our analysis.

Next, ITO and Au antennas on BST substrates were analyzed. The dielectric constant for BST was taken to be 300. The conductivity for the ITO is taken to be $1.3 \times 10^4$ and the dielectric constant to be 3.37+0.01j. Antennas with different gap separation, metal thickness, width, and length were simulated to optimize the antenna's dimensions. The results of the simulation are shown if Figs. 5, 6, 7, 8, 9, 10, 11, and 12. The general behavior of the Au/BST and ITO/BST antennas is the same, but the ITO antennas have higher outputs.

The Au and ITO antennas were simulated using gap spacing of 100, 200, 300, 400, and 500 nm. The results show that the electric field increases with the decrease of the gap separation. Fig. 5 shows the electric field as a function of the gap separation for Au/BST antenna and Fig. 6 shows the same results for the ITO/BST antenna. The two figures show also that the resonant frequency does not change with the gap separation. This is because the resonant frequency is independent of this gap dimension. The highest electrical field was found to be at 100 nm and the lowest at 500 nm. The max electric field strength can be obtained at the minimum gap dimension, which is determined by the minimum line-width of the fabrication process used. The increase in the field with the decrease of the gap can be attributed to the fact that the electric field is the voltage over the gap distance; and consequently, as the gap decreases, the electric field strength increases.

The antennas' performance was then analyzed for different metal thicknesses, but with the original width, length, and gap dimensions. The simulation was done using thicknesses of 0.2, 0.5, 1, 3, and 5 μm. The results of the study are shown in Fig. 7 for the Au/BST antenna and Fig. 8 for the ITO/BST antenna. For both antennas, a higher thickness resulted in a higher output field. The resonant frequency of Au and ITO antennas are the same, indicating that this frequency is independent of the metal thickness. Therefore, the increase of the electric field with the thickness can be attributed to the increase of the metal conductance.

To investigate the effect of the Skin depth on the antenna with smaller thickness, the skin depth was calculated using the equation:

$$\delta_s = \sqrt{\frac{1}{\pi * f * \mu * \sigma}} \qquad (3)$$

where, $\delta_s$ is skin depth, f is frequency, μ is permeability of material and σ is conductivity. The skin depth was found to be 70 nm for 1 THz frequency. The analysis shows us that even with 1um thick the Skin depth is only 7 % of the thickness. It was, then, concluded that skin depth for the metals has no effect, and that it is the metal conductance that has the dominant effect on the antenna's performance.

The effect of the width dimension on the antenna's response was investigated next. The antennas were simulated with widths of 2, 4, 9, 10, and 12 μm. The results for the Au antenna are shown in Fig. 9, while the results for the ITO antenna are shown in Fig. 10. The two figures show that the electric field strength is invariant with the width. This can be attributed to the fact that the applied electromagnetic wave has the magnetic field parallel to the width and the electric field parallel to the length. This means the width dimension affects the magnetic field only and the length dimension affects the electric field only. Therefore, the Electric Field stays constant with the width variation. If the magnetic field is simulated, the results are expected to be a function of the antenna's width.

The impact of the length on the output electric field was also investigated. The antennas were simulated using lengths of 25 μm, 20 μm, 16 μm, 12 μm, and 8 μm. The simulation results for the Au/BST antennas show that the resonant frequency changes with the change in the length. As can be seen from Fig. 11, the Au/BST antenna resonate at 0.54 THz, 1.443 THz, and 2.165 THz for the lengths of 16, 12, and 8 μm. For lengths of 20 and 25 μm, the primary frequency falls outside the simulation range. For the ITO/BST antennas, the resonant frequency changes with the change in the length too. As can be seen from Fig. 12, the primary resonant frequency is the same as those of Au/BST antenna. This is because, the resonant frequency depends on the substrate properties, not on the metal.

In an antenna, there may exist multiple peaks of the electric field, since these antennas resonate at frequencies $f_0/n$, $f_0$. n $f_0$, where n is an integer and $f_0$ is the fundamental frequency. The resonant frequency in this case can be obtained from (4).

$$f_0 = \frac{c}{2 * L * \sqrt{\varepsilon_{eff}}} \quad (4)$$

Where, $c$ is the speed of light, $L$ is the length of the antenna and $\varepsilon_{eff}$ is the effective dielectric constant at $f_0$. The antennas resonate, not only at the fundamental frequency, but also at the multiples or fraction of the fundamental frequencies because of the periodic nature in the geometry. Our results are in agreement with this statement, as shown in Fig. 11 and Fig. 12. For example, Au/BST and ITO/BST antennas, with L= 16 µm, resonate at a fundamental frequency of 0.54 THz and a secondary frequency of $2f_0$ = 1.08 THz. For L = 12 µm, the antenna does not resonate at $f_0$ = 0.72 THz, but it does resonate at 2 $f_0$ of about 1.44 THz, which is a secondary frequency.

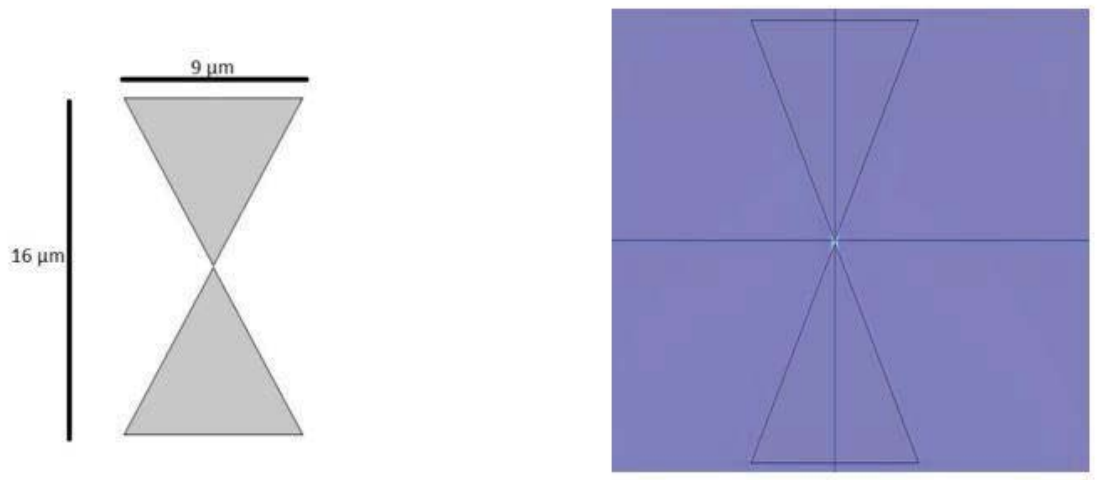

Fig. 1. Geometry and COMSOL simulation of Bow-Tie Antenna used in this Study: width is 9 µm, length is 16 µm, gap is 100 nm, and thickness 1 µm.

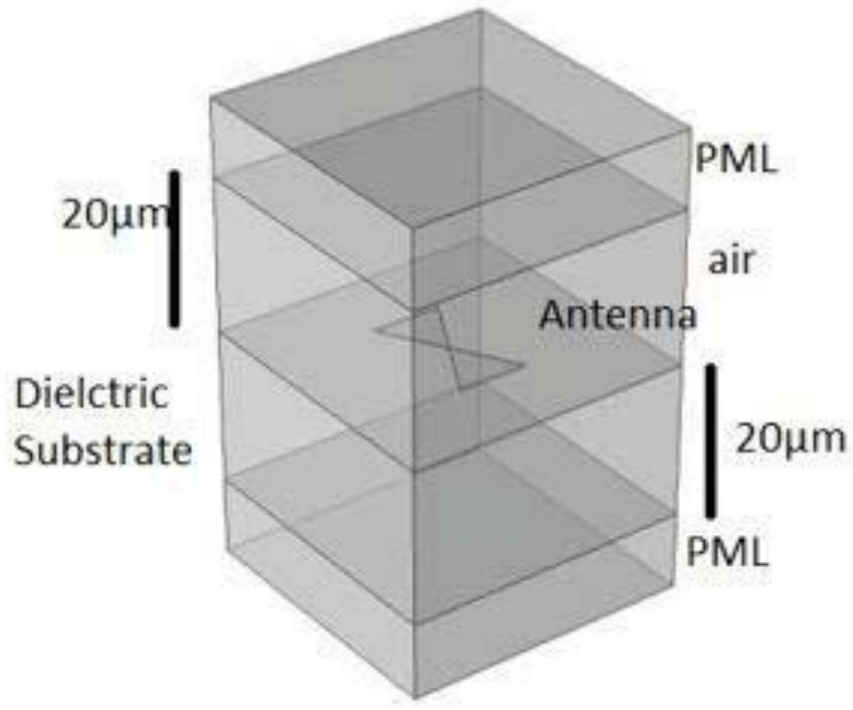

Fig. 2. A 3-D view of Simulation domain in COMSOL with a Perfectly Matched Layer (PML) on the top and at the bottom

In conclusion, this study investigated the performance of Au/BST and ITO/BST antennas. The study shows how the dimensions can affect the performance of the antenna. It also shows that, using an array of antennas with different lengths, a band pass antenna with different paasing with can be built to allow the detection of specific range of frequencies.

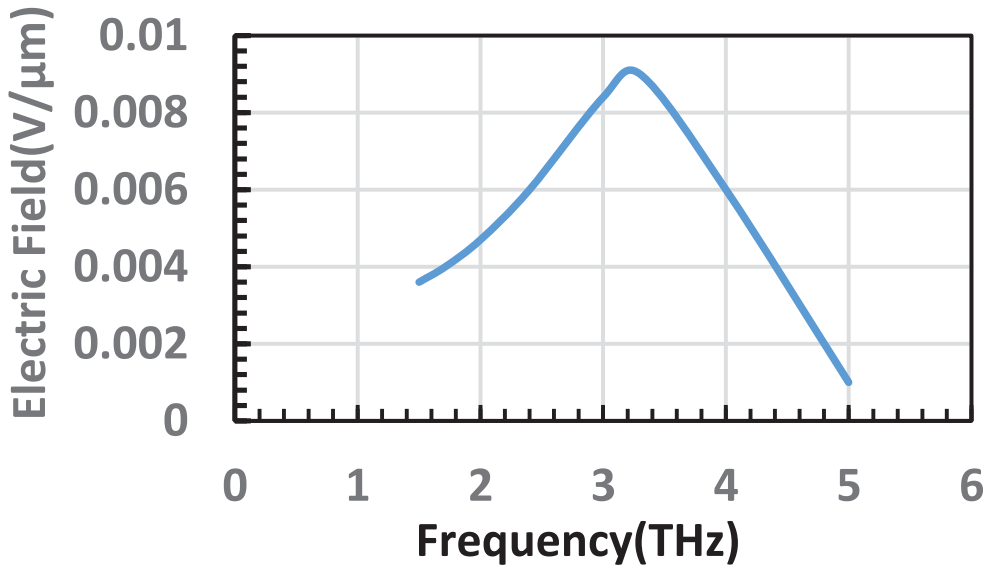

Fig. 3. Output of Au/Silica Antenna reported in [5]

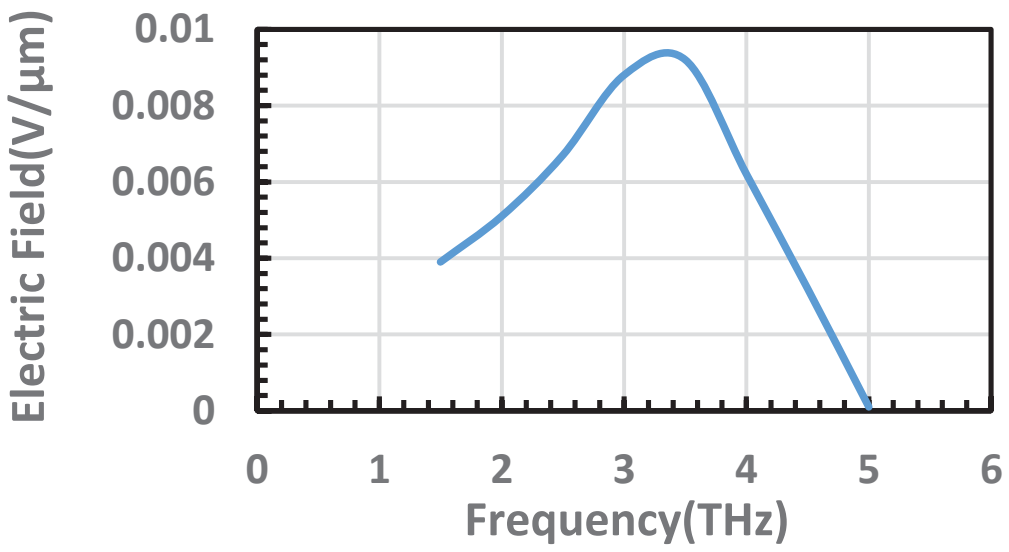

Fig. 4. Output of Au/Silica Antenna from this study

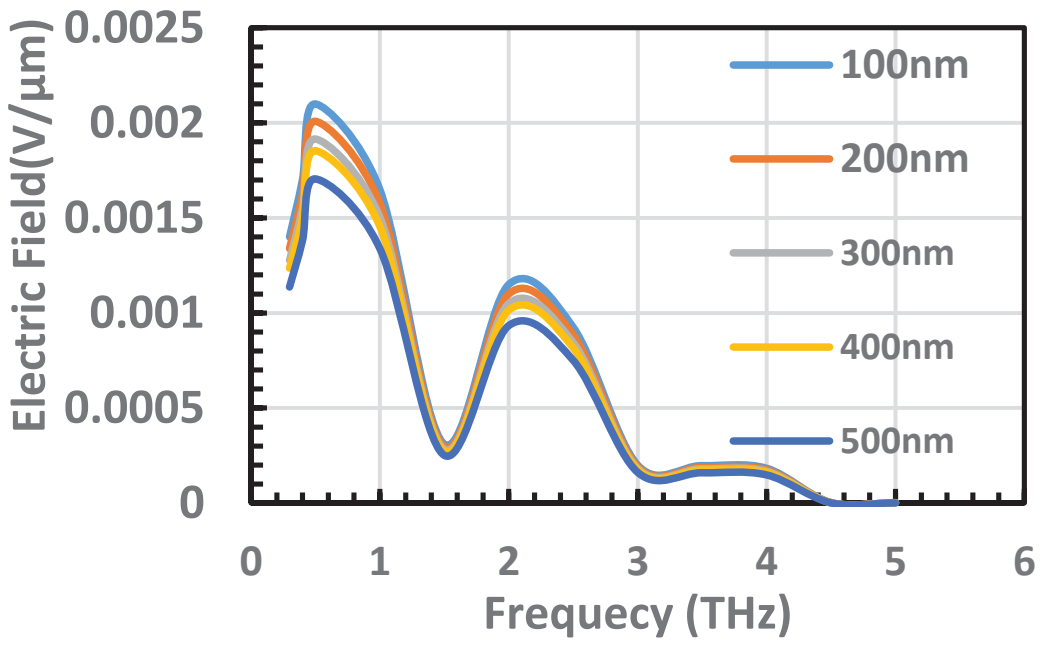

Fig. 5. Electric Field from Au/BST Antenna optimized for gap between antenna elements

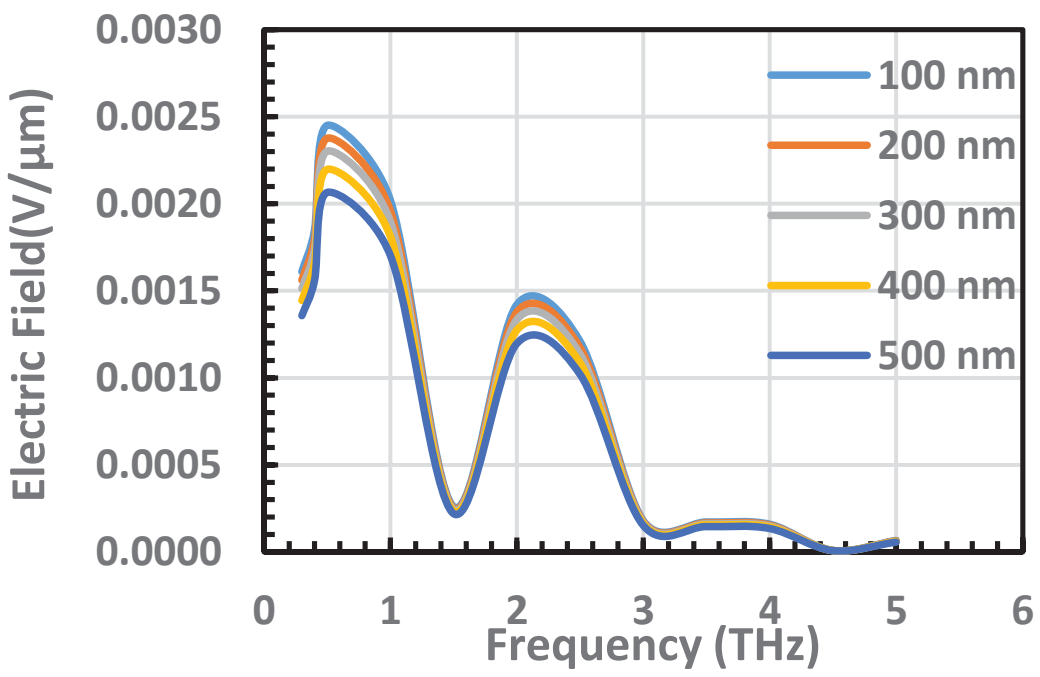

Fig. 6. Electric Field from ITO/BST Antenna optimized for gap between antenna elements

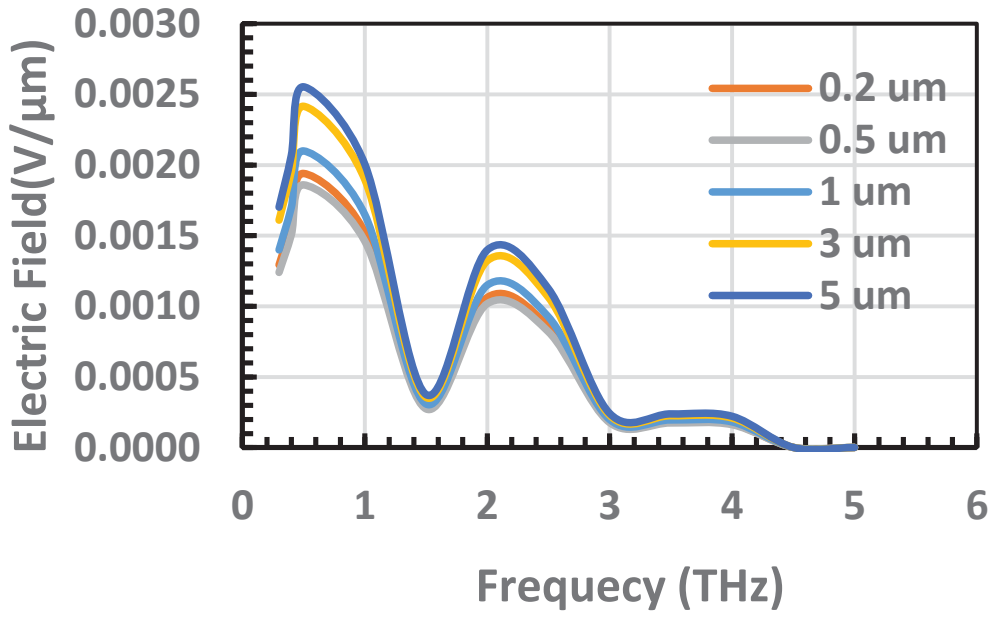

Fig. 7. Electric Field from Au/BST Antenna optimized for thickness

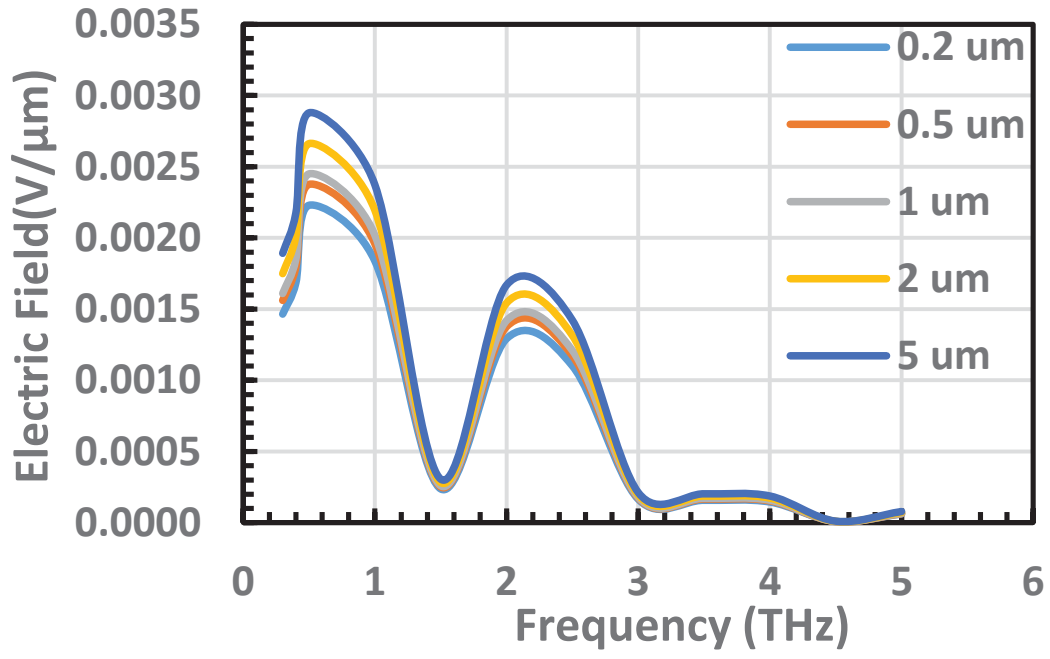

Fig. 8. Electric Field from ITO/BST Antenna optimized for thickness

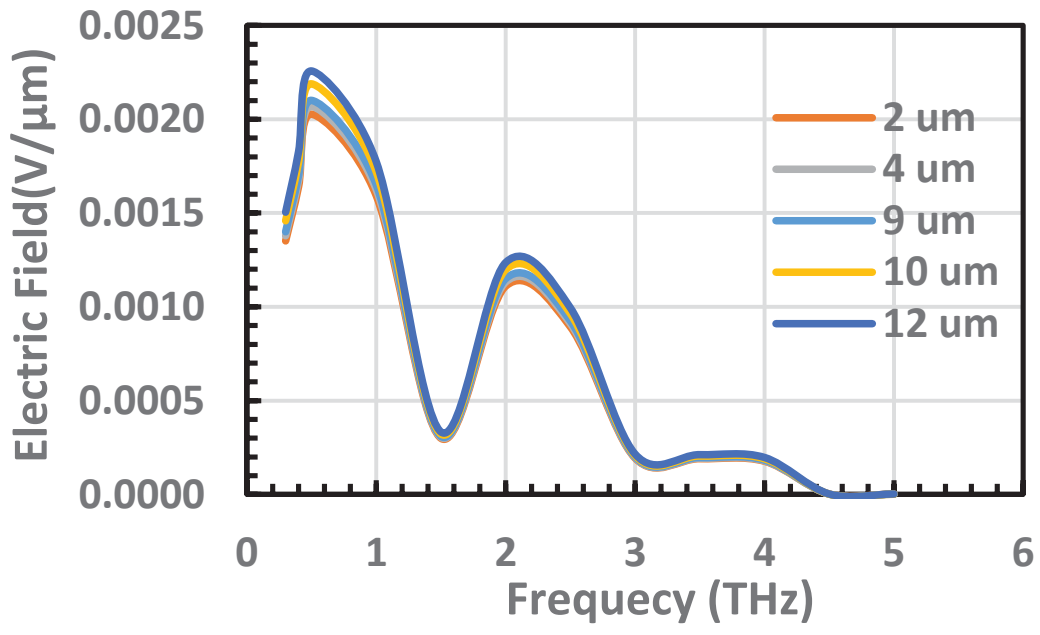

Fig. 9. Electric Field from Au/BST Antenna optimized for width

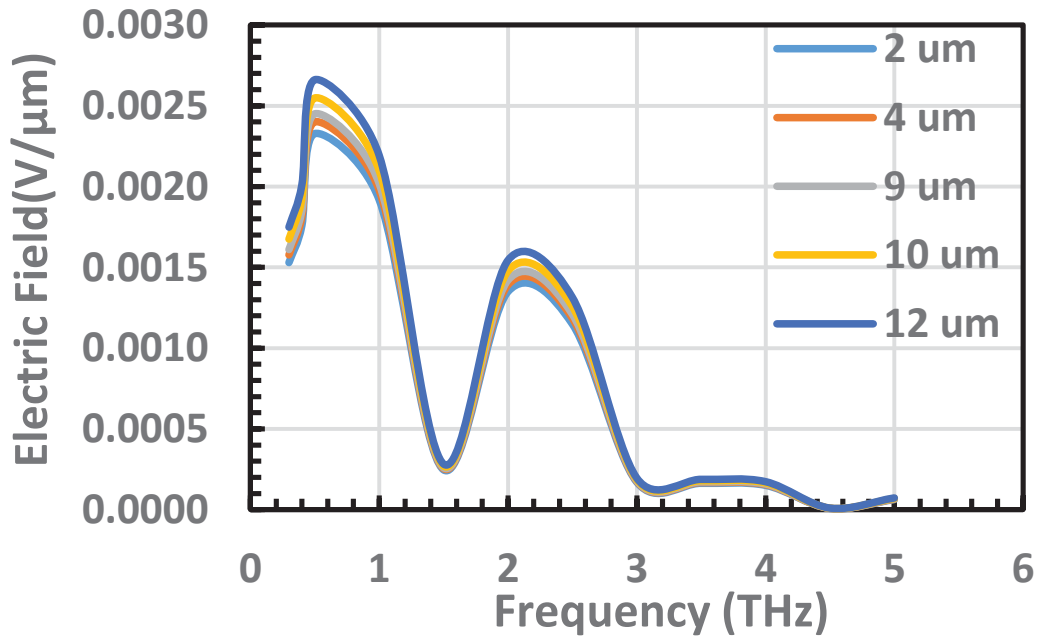

Fig. 10. Electric Field from ITO/BST Antenna optimized for width

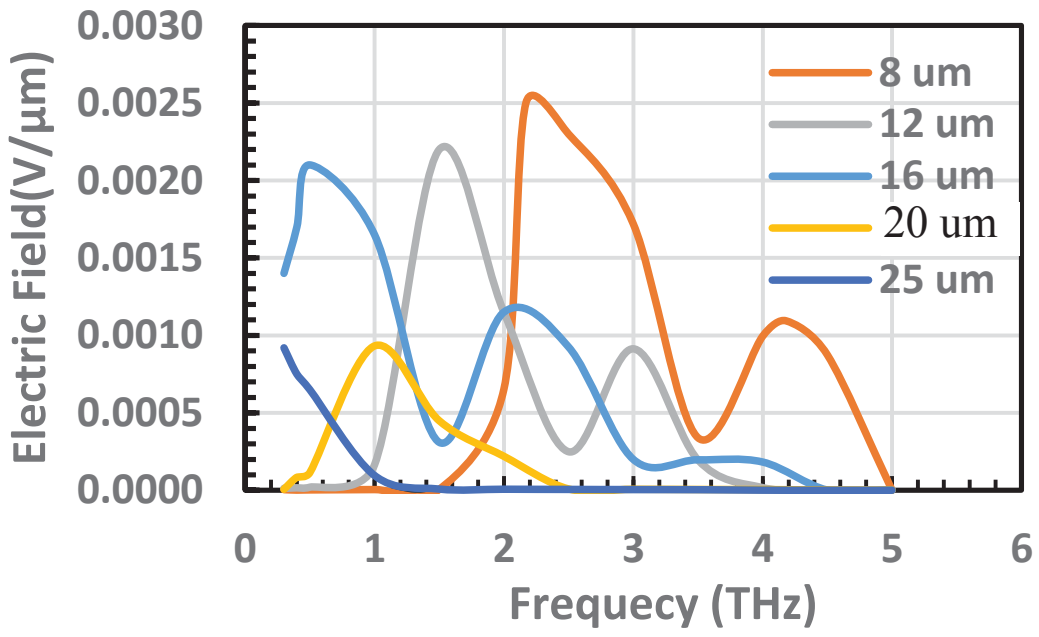

Fig.11. Electric Field from Au/BST Antenna optimized for length

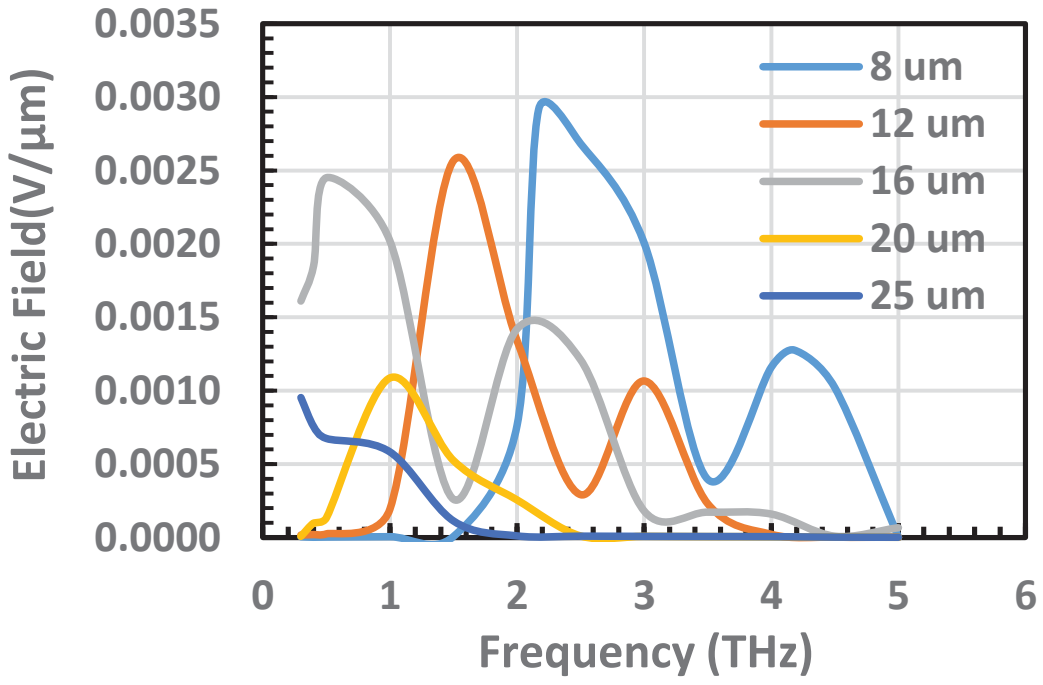

Fig. 12. Electric Field from ITO/BST Antenna optimized for length